\RequirePackage{fix-cm}  
\documentclass[aps,pra,showpacs,amsmath,amssymb,amsfonts,superscriptaddress,lengthcheck,longbibliography, twocolumn]{revtex4-1}
\usepackage{graphicx}
\usepackage{subfigure}
\usepackage{amsthm}
\usepackage{verbatim}
\usepackage{dcolumn}
\usepackage{bm}
\usepackage{epsf}
\usepackage{color}
\usepackage[colorlinks=true,citecolor=blue,linkcolor=blue,urlcolor=blue]{hyperref}%
\newcommand{\bra}[1]{\left\langle #1\right|}
\newcommand{\ket}[1]{\left|#1\right\rangle}

\newcommand{\bla}{bla\\bla\\bla\\bla\\bla}

\newcommand{\bfm}{\boldsymbol}

\begin{document}
%
\title{Quantum neural networks to simulate many-body quantum systems}
\author{Bart\l{}omiej Gardas}
\email{bartek.gardas@gmail.com}
\affiliation{Theoretical Division, LANL, Los Alamos, New Mexico 87545, USA}             
\affiliation{Institute of Physics, University of Silesia, 40-007 Katowice, Poland}     
\affiliation{Jagiellonian University, Marian Smoluchowski Institute of Physics, \\ \L{}ojasiewicza 11, 30-348 Krak\'ow, Poland}
\author{Marek M. Rams}
\author{Jacek Dziarmaga}
\affiliation{Jagiellonian University, Marian Smoluchowski Institute of Physics, \\ \L{}ojasiewicza 11, 30-348 Krak\'ow, Poland}

\date{\today}
\begin{abstract}	
We conduct experimental simulations of many body quantum systems using a \emph{hybrid} classical-quantum algorithm.
In our setup, the wave function of the transverse field quantum Ising model is represented by a restricted Boltzmann machine. 
This neural network is then trained using variational Monte Carlo assisted by a D-Wave quantum sampler to find
the ground state energy. Our results clearly demonstrate that already the first generation of quantum computers
can be harnessed to tackle non-trivial problems concerning physics of many body quantum systems.
\end{abstract}

\maketitle

\section{Introduction} 
Building a universal quantum computer is a holy grail of modern sciences~\cite{Cong17}. Such machine  offers \emph{necessary} 
capabilities allowing one to simulate \emph{highly} entangled quantum systems~\cite{Petruccione16,Adolfo17}. In contrast, universal 
Turing machine~\cite{Couder16} realizing classical computation can only simulate \emph{slightly} entangled quantum states~\cite{Vidal03}.  

While quantum supremacy is yet to be demonstrated~\cite{Terhal2018}, many important properties of quantum systems can be captured by artificial intelligence~\cite{Petruccione14} and neural networks in particular~\cite{Deng17,Gao17}. The so called quantum neural states provide a novel ansatz to represent the wave function of a many body quantum system~\cite{Carleo18,Cirac18}. Such neural networks can be \emph{taught} using various techniques -- most notably the variational Monte Carlo~\cite{Sorella98,Carlson15,Foulkes01}.
In general, however, sampling the state space, which is the key ingredient of all Monte Carlo methods, cannot be executed efficiently by any classical algorithm~\cite{Troyer04,Nielsen2010}. Hence, there exist natural limitations to any classical algorithm that aims to teach the network about quantum systems. 

It is well known that these limitations can be broken by harnessing the power of a quantum sampler~\cite{Sinitsyn18,Denchev16}. It is needless to say that the existing annealers are far from perfect~\cite{Smolin14,Albash2015,konz_uncertain_2018}. Nevertheless, they can be turned into quantum samplers rather easily. This provides an ideal playground for testing a new generation of \emph{hybrid} classical-quantum algorithms~\cite{Wittek17}.  

In this work, we investigate to what extent such algorithms can run on the existing hardware~\cite{Lloyd17}. Our purpose is to demonstrate that already the first generation of quantum computers can in fact \emph{assist} in simulations of simple yet non-trivial many body quantum systems. Similar conceptual idea has been applied \emph{very} recently to investigate quantum phase transition in many body systems~\cite{king_observation_2018,harris_phase_2018}, risk analysis~\cite{woerner_risk_2018} and quantum circuits diagonalizing 
{\it small} quantum Hamiltonians~\cite{kandala_hardware-efficient_2017,lierta_IBM_2018}.
In our setup, however, the wave function of the transverse quantum Ising model is represented by a restricted Boltzmann machine~\cite{Carleo17}. This neural network is then trained in an \emph{unsupervised} manner to find the ground state energy. The learning process is assisted by a D-Wave chip as explained below.
\section{Preliminaries}
\subsection{Quantum neural states}
We begin by writing a many body quantum state $\ket{\psi}$ using restricted Boltzmann
machine (RBM) as a wave function ansatz~\cite{Carleo17,Chen18}:
\begin{equation}
\label{eq:rbm}
\ket{\Phi} = \sum_{\bfm v} \Phi(\bfm v)\ket{\bfm v}, \quad
\Phi(\bfm v) = \sum_{\bfm h } e^{-\phi(\bfm v,\bfm h) }.
\end{equation}
Here $\bfm v = \left(v_1,\dots, v_N\right)$ is a collection of physical degrees of freedom called \emph{visible} neurons, 
\begin{equation}
\label{eq:rmbe}
\phi(\bfm v,\bfm h) = \bfm a \cdot \bfm v 
                                + \bfm b \cdot \bfm h 
                                + \bfm h \cdot \bfm W \cdot \bfm v
\end{equation}
and $\bfm h = \left(h_1,\dots, h_M\right)$ are \emph{hidden} neurons, see Fig.~\ref{fig:RMB}. This network is fully 
specified by the weights $\bfm a$, $\bfm b$, $\bfm W$ which are determined during the learning stage. Surprisingly, 
$M = \alpha N$ for moderate $\alpha$, say $<4$, is often sufficient to accurately calculate the ground state properties
of many important physical systems~\cite{Carleo17}. 
 
Our objective here is to train the quantum neural state using a D-Wave annealer to find the ground 
state energy $E$ of the transverse field quantum Ising model~\cite{Dziarmaga05,Zurek05,Gasenzer18},
\begin{equation} 
\label{eq:ising}
H = -h\sum_{i}\hat{\tau}^x_i-\sum_{\langle i, j\rangle}\hat{\tau}^z_i\hat{\tau}^z_{j}.
\end{equation}
Above, $\langle i, j\rangle$ denotes nearest neighbors and $\tau^x_i$, $\tau^x_i$ are the standard Pauli spin 
operators~\footnote{We reserve symbols $\sigma^x_i$, $\sigma_i^z$ for the annealer Hamiltonian.}.
Periodic boundary conditions are assumed. We consider both $1$D and $2$D lattices.
In the former case, the ground state energy can be calculated exactly~\cite{Lieb61}. In the latter, we use density
matrix renormalization group algorithm~\cite{Schollwock05,Wall18} to obtain its sufficient approximation that 
will serve as a reference value. This allows us to assess the robustness of our method and at the same time 
benchmark the annealer~\cite{Gardas18}. To perform a quantum sampling we use both the newest $2000$Q 
chip and its predecessor DW$2$X~\cite{Gardas17}.

Henceforward, we also assume that all weights $\bfm a$, $\bfm b$, $\bfm W$ are \emph{real}. Thus, 
instead of Eq.~(\ref{eq:rbm}), one can use the following ansatz to represent the ground state of the Ising model~(\ref{eq:ising}):
\begin{equation}
   \label{eq:sa}
   \Psi(\bfm v) = \sqrt{ \sum_{\bfm h } e^{-\phi(\bfm v, \bfm h) } } = \sqrt{\Phi(\bfm v) }.
\end{equation}
As before, $\phi(\bfm v, \bfm h)$ is given by Eq.~(\ref{eq:rmbe}). 
This way, the \emph{quantum} probability distribution,
\begin{equation}
\label{eq:qdist}
 \rho(\bfm v) = \frac{|\Psi(\bfm v)|^2  }{  \sum_{\bfm v^{\prime}} |\Psi(\bfm v^{\prime})|^2},
\end{equation}
can be represented by a RBM and as such can be sampled using a quantum annealer. 
This is the key insight into all conceptual ideas we outline in this article.
\subsection{Adiabatic quantum computing} 
\begin{figure}
	\includegraphics[width=\columnwidth]{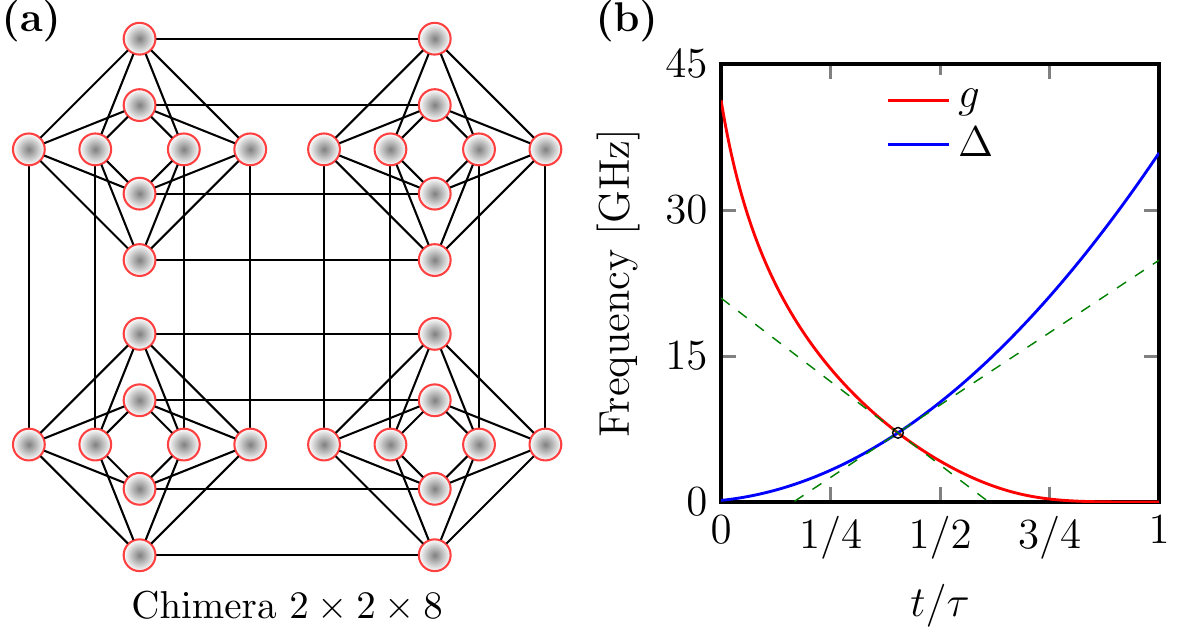}
	\caption{\emph{D-Wave processor specification}.
	{\bf (a)} An example of the chimera architecture comprising of a $2 \times 2$ grid consisting of
	clusters (units cells) of $8$ qubits each.
	{\bf (b)} A typical annealing schedule, where the annealing time reads $\tau$.}
	\label{fig:DW}
\end{figure}
During quantum annealing a many-body system is to be evolved from the ground state of a problem Hamiltonian 
$\mathcal{H}_0$ to the ground state of a final Hamiltonian $\mathcal{H}(\tau)$ that encodes the solution for the 
problem of interest~\cite{Nishimori98,Farhi00,Aharonovetal}. The dynamics of a D-Wave chip is governed by 
the time-dependent Hamiltonain~\cite{Lanting14,Sinitsyn18}
\begin{equation} 
\begin{split}
\mathcal{H}(t) / h &= -g(t)\sum_{i\in\mathcal{V}}\hat{\sigma}_i^x - \Delta(t) \mathcal{H}_0, \\
\mathcal{H}_0 &= \sum_{i\in\mathcal{V}}B_i \hat{\sigma}_i^z + \sum_{(i,j)\in \mathcal{E}} J_{ij}\hat{\sigma}_i^z \hat{\sigma}_{j}^z,
\label{eq:H0}
\end{split}
\end{equation}
where $\hat{\sigma}_i^x$ and $\hat{\sigma}_i^z$ are the Pauli spin operators. 
Here, $\mathcal{H}_0$ is defined on a chimera graph $\mathcal{G}=(\mathcal{V}, \mathcal{E})$~\cite{Mniszewski17}, see 
Fig.~\ref{fig:DW}{\color{blue} (a)}.
\emph{Dimensionless} couplers $J_{ij}$ and biases $B_i$ can be controlled by the users, but only within a predefined range. 
For example, on the $2000$Q device we have $|J_{ij}|\le 1$ and $|B_i|\le 2$. 

Presumably, the Hamiltonian $\mathcal{H}(t)$ varies slowly while $\Delta(t)$ is changed from initial $\Delta(0)\approx 0$ to large final $\Delta(\tau)$, 
wheres $g(t)$ varies from large initial $g(0)$ to final $g(\tau)\approx 0$ [cf. Fig.~\ref{fig:DW}{\color{blue} (b)}]. As a result, under optimal conditions, 
the system remains in its ground state.
Thus, the desired solution -- encoded in eigenvalues, $\sigma_i$, of $\hat{\sigma}_i^z$ -- can be extracted through a measurement of 
the final state. It is worth emphasizing that with D-Wave annelers one can only carry a suitable measurement in the computational  
$z$-basis ($\ket{\uparrow}$, $\ket{\downarrow}$). This essentially excludes any possibility to measure the ground state energy of the 
transverse Ising model~(\ref{eq:ising}) {\it directly}, even if measurements for intermediate values of $\Delta$ and $g$ were available.

Furthermore, since no real hardware is completely isolated from its environment, the outcome of such experiment  will be 
distributed according to some \emph{temperature}-dependent probability distribution $p(\bfm \sigma)$~\cite{Yarkoni16}. 
Relaxing the annealer to the equilibrium, one can approximate $p(\bfm \sigma)$ by the \emph{classical} Boltzmann distribution~\cite{Benedetti17},
\begin{equation} 
\label{eq:drbm}
p(\bfm \sigma) \sim e^{ -\beta E_{\tau}(\bfm \sigma) },
\end{equation}
where the energy function reads
\begin{equation} 
\label{eq:Hf}
E_{\tau}(\bfm \sigma) = -\sum_{i\in\mathcal{V}}B_i\sigma_i-\sum_{(i,j)\in \mathcal{E}} J_{ij}\sigma_i\sigma_{j}.
\end{equation}
Time to complete the annealing cycle is denoted by $\tau$. The \emph{effective} inverse temperature,
$\beta = h\Delta(\tau^{*})\beta_{\rm chip}/k_B$,  is affected by many factors, including the specific values of the 
control parameters $J_{ij}$ and  $B_i$~\cite{Amin15}. Here, $\tau^*\le \tau$ is the so called freeze-out time and
$1/\beta_{\rm chip}$ denotes the chip's operational temperature~\footnote{To sample form a \emph{classical} 
distribution $g(\tau^*) \ll \Delta(\tau^*)$.}. 
Note,  $\beta $ is \emph{a priori} unknown and
it can only be determined on a case-by-case basis~\cite{Benedetti16}. In this work, however, we do not attempt 
to estimate the function $\beta(J_{ij},B_i,\tau^*)$. We rather try to modify the sampling algorithm to account for its 
possible variation with the values of the parameters. 
 
In an annealer with the sufficient connectivity between qubits, there would be a one-to-one mapping between the set of $\sigma_i$ and the two sets of visible 
and hidden neurons, $\bfm \sigma = [\bfm v, \bfm h]$. Accordingly, every nonzero $J_{ij}$ would be identified 
with a $W_{ij}/\beta$ between the visible and hidden neuron and the biases $\bfm B = [\bfm a, \bfm b]/\beta$. In practice,
\begin{equation}
J_{ij} = W_{ij} /\beta_{\rm x},
\end{equation}
where $\beta_{\rm x}$ is an \emph{estimation} of the inverse temperature $\beta$.
\section{Unsupervised learning} 
\subsection{Variational Monte Carlo} 
Training neural networks can be tedious. Moreover, due to its topology, RBMs may be highly susceptible to small changes of the variational parameters. Their adjustments can further propagate throughout the network causing even larger changes of the wave function. To mitigate these problems we use stochastic reconfiguration, a method that is widely used in the variational Monte Carlo~\cite{Sorella98}. At each iteration the network weights, $\bfm w = [\bfm a,\bfm b,\bfm W]$, 
are refined according to
\begin{equation} 
\label{eq:update}
\bfm w_{k+1} = \bfm w_{k} - \gamma_k \bfm x_k, \quad \bfm{S}^{(k)} \bfm x_k =  \bfm{F}^{(k)},
\end{equation}
where a non-negative definite covariance matrix reads 
\begin{equation} 
\label{eq:cov}
 \bfm{S}_{ij} = \langle\langle D_i^* D_j \rangle \rangle_{\rho} - 
                      \langle\langle D_i^*  \rangle\rangle \langle\langle D_j  \rangle\rangle_{\rho},
\end{equation}
and the so called \emph{forces} are given by
 \begin{equation} 
 \label{eq:force}
 \bfm{F}_j = \langle\langle E D_j^* \rangle \rangle_{\rho} - 
                    \langle\langle E  \rangle\rangle_{\rho} \langle\langle D_j^*  \rangle\rangle_{\rho}.
 \end{equation}
Double brackets $\langle\langle\cdot\rangle \rangle_{\rho}$ indicate averages with respect to the distribution in Eq.~(\ref{eq:qdist}), $\gamma_k$ is the learning rate. Finally, 
\begin{equation} 
\label{eq:deriv}
D_i = \frac{1}{\Psi(\bfm v)} \frac{\partial}{\partial w_i} \Psi(\bfm v), \quad 
  E_{\rm loc} = \frac{\bra{\bfm v} H \ket{\Psi}}{\Psi(\bfm v)},
\end{equation} 
denote the gradients and local energy, respectively~\cite{Carleo17}. 
\subsection{Sampling} 
Usually, the importance sampling is performed using Metropolis-Hastings algorithm~\cite{Calderhead14}. In this work we employ the newest generations of D-Wave samplers to calculate covariance matrix~(\ref{eq:cov}) and forces~(\ref{eq:force}) at each iteration~\cite{Amin16}. The remaining part of the algorithm is executed on a classical processing unit: CPU or GPU~\cite{GPU2018}. 

\begin{figure}[t]
	\includegraphics[width=\columnwidth]{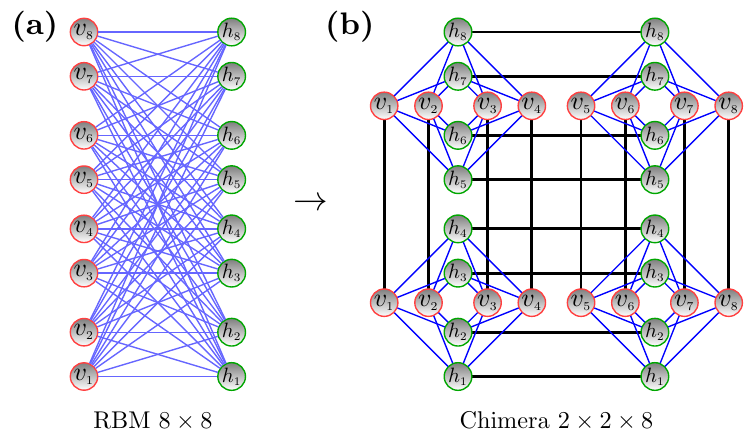}
	\caption{\emph{Restricted Boltzmann machine embedded on D-Wave}.
	{\bf (a)} Graphical representation of the neural network~(\ref{eq:rmbe}) with
		$8$ visible and $8$ hidden neurons.
	   {\bf (b)} The same network embedded on the chimera graph $\mathcal{G} = (\mathcal{E}, \mathcal{V})$ of size $2\times 2 \times 8$.
	    Strong ferromagnetic couplings (thick lines)
	    ``glue'' qubits in different unit cells to represent single neurons.
       }
	\label{fig:RMB}
\end{figure}

To this end, we first rewrite the ansatz~(\ref{eq:sa}) as 
\begin{equation}
\Psi(\bfm v) = e^{\bfm a \cdot \bfm v / 2} \, \left[ \prod_{j=1}^M 2 \cosh\left(b_j + \bfm W_{j} \cdot \bfm v\right)\right]^{1/2},
\label{psis}
\end{equation}
where we explicitly traced out hidden variables. This is possible due to the lack of intra-layer interactions between  hidden
neurons [cf.~Fig~\ref{fig:RMB}{\color{blue} (b)}]. Now, all derivatives in Eq.~(\ref{eq:deriv}) can be expressed using visible neurons only~\cite{Carleo17},
\begin{equation}
   \frac{1}{\Psi(\bfm v)}\frac{\partial}{\partial p} \Psi(\bfm v)  = \frac{1}{2} \cdot
      \begin{cases}
           v_i, &  p = a_i \\
            \tanh\left(\theta_j\right) & p = b_j \\
           v_i  \tanh\left(\theta_j\right) & p = W_{ij}
    \end{cases}
\end{equation}
where we introduced $\theta_j = b_j + \bfm W_{j} \cdot \bfm v$. Similarly, the local energy can be simplified to take the form
\begin{equation}
E_{\rm loc}  = -h \sum_{i} \frac{\Psi(\bar{\bfm  v}_i)}{\Psi(\bfm v)} - \sum_{\langle i, j\rangle} v_i v_{j},
\end{equation}
where $\bar{\bfm  v}_i$ denotes a vector ${\bfm v}$ with $i$-th spin flipped and $\Psi$ is given by Eq.~(\ref{psis}). Finally, to compute 
$\langle\langle\cdot\rangle \rangle_{\rho}$ using samples gathered from a quantum annealer, 
we note that
\begin{equation}
\begin{split}
\langle\langle f \rangle \rangle_{\rho} &= \sum_{\bfm v} \rho(\bfm v) f(\bfm v)  
\approx  \sum_{\bfm v, \bfm h} p(\bfm v, \bfm h) f(\bfm v) \\
&= \sum_{\bfm \sigma } p(\bfm \sigma)  f(\bfm \sigma) 
\approx  \frac{1}{N_{\rm s}} \sum_{i=1}^{N_{\rm s}} f(\bfm \sigma_i),
\end{split} 
\end{equation}
where the \emph{bare} output of the D-Wave annealer, $\bfm \sigma = [\bfm v, \bfm h]$, encodes both hidden and visible neurons. Here, $N_{\rm s}$
is the number of samples. The first approximation is true under a proper embedding as long as $p(\bfm v, \bfm h)$ is \emph{close} to the Boltzmann distribution. The second one holds for sufficiently large $N_{\rm s}$ (in practice $N_{\rm s} \sim 10^4$) provided that all samples $\bfm \sigma_i$ are distributed according to $p(\bfm \sigma_i)$. Note, one quantum annealing corresponds to one sample. This means that gathering
samples requires running the annealer over and over again. However, in practice this is executed in ``chunks'' and it is very fast ($\mu$s - ms depending on the annealing time $\tau$) and efficient.

A possible advantage of using a D-Wave computer is that it can sample both visible and hidden neurons simultaneously. In principle this allows to calculate the gradients directly using the ansatz in Eq.~\eqref{eq:sa}, even try to extend it to \emph{deep} Boltzmann machine~\cite{tramel_deterministic_2018}. We leave this for future investigations.

 \begin{figure}[t!]
 	\includegraphics[width=\columnwidth]{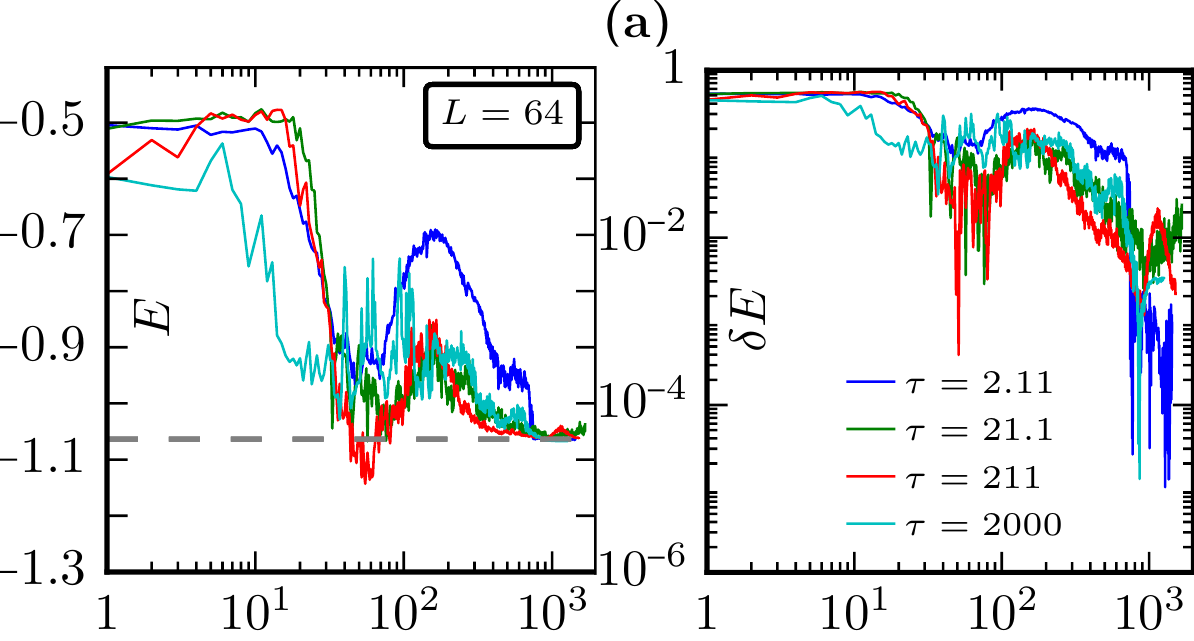}
    \includegraphics[width=\columnwidth]{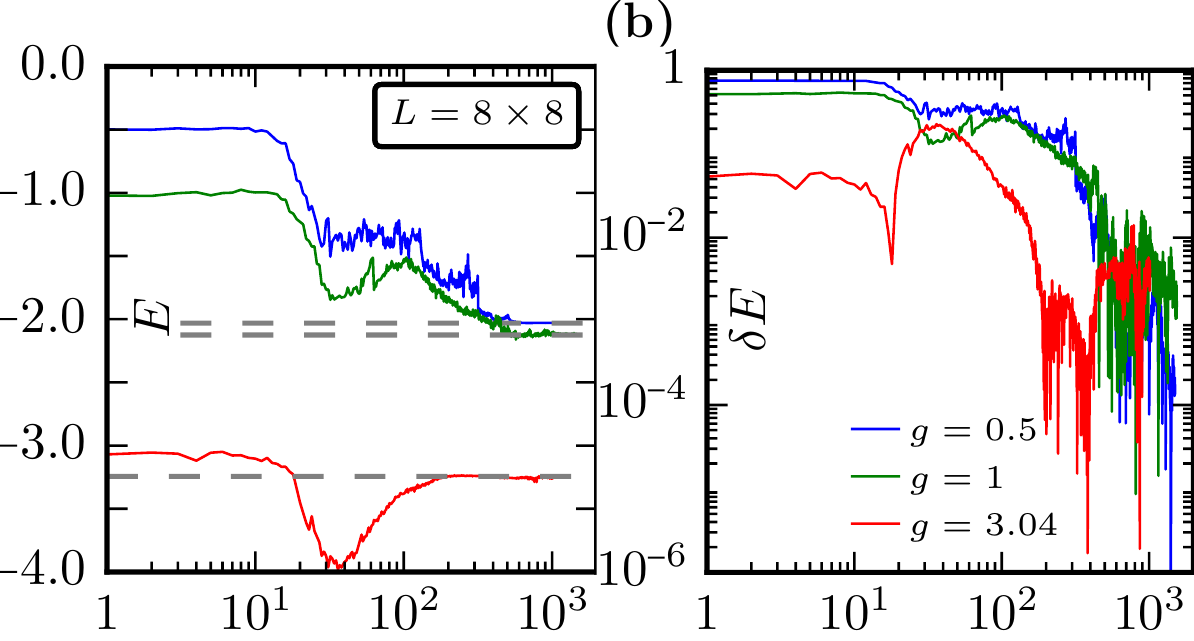}	 
 	\includegraphics[width=\columnwidth]{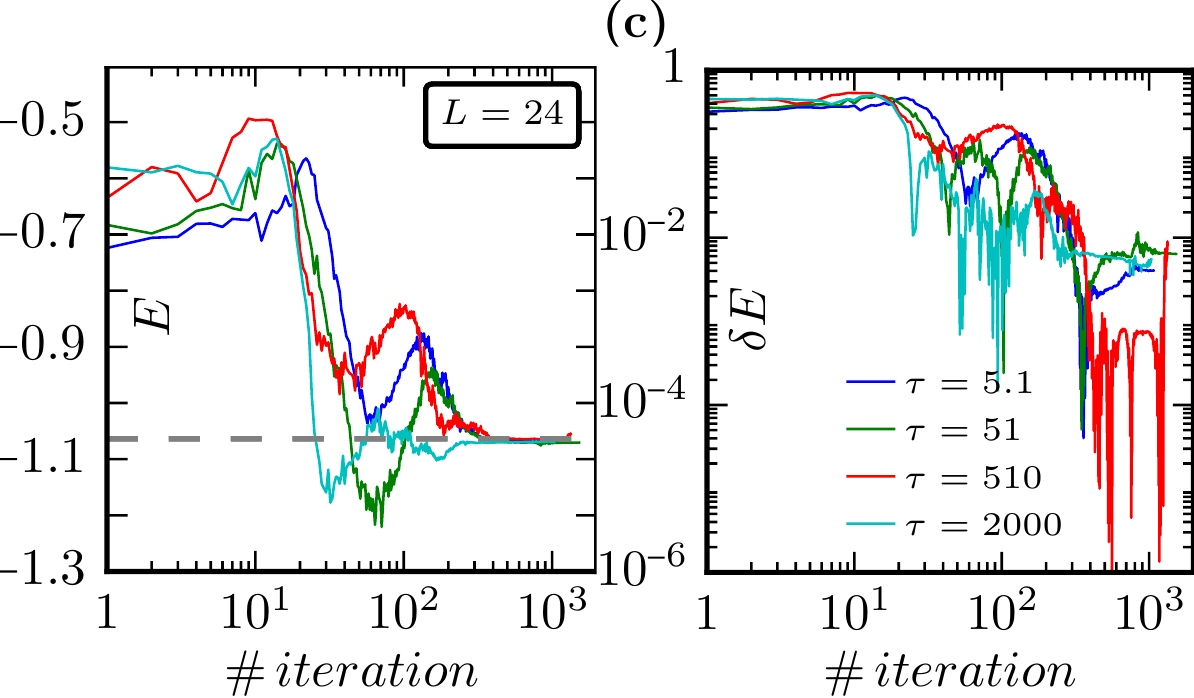}	
 	\vspace{-0.0cm}
 	\caption{
 		The ground state energy per spin $E$ for the transverse field quantum Ising models~(\ref{eq:ising}) in 1D and 2D. Dashed lines indicate the exact values. Sampling was executed using D-Wave annealers: $2000$Q in (a) and (b) and DW$2$X in (c).  $\delta E$ shows the relative energy error reached. We set $h=0.5$ for 1D system in panels (a) and (c), and $h=0.5$, $1$ and $3.044\simeq h_c$ for 2D system in (b), where $\tau=20$.
		The annealing time $\tau$ is measured in $\mu s$. Besides, $\alpha=1$, $\gamma=0.2$ and $N_{\rm s}=10^4$. 
 	}
 	\label{fig:energy} 
 \end{figure}
 
 \begin{figure}[t!]
 	\begin{center}
 		\includegraphics[width=\columnwidth]{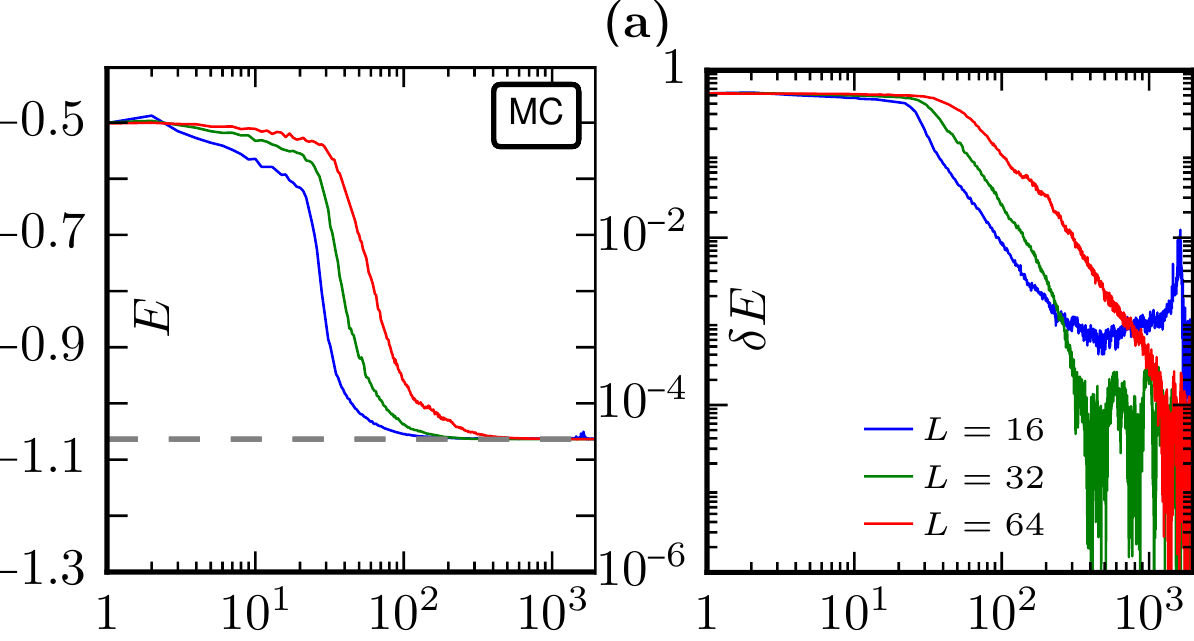}
 		\includegraphics[width=\columnwidth]{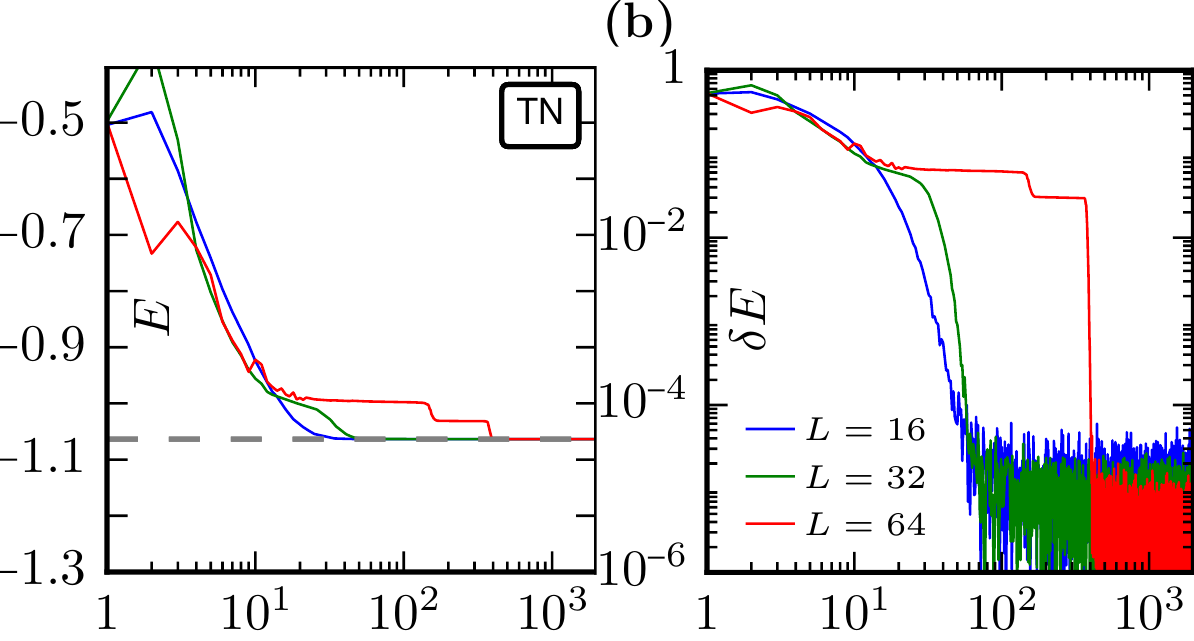}
 	 	\includegraphics[width=\columnwidth]{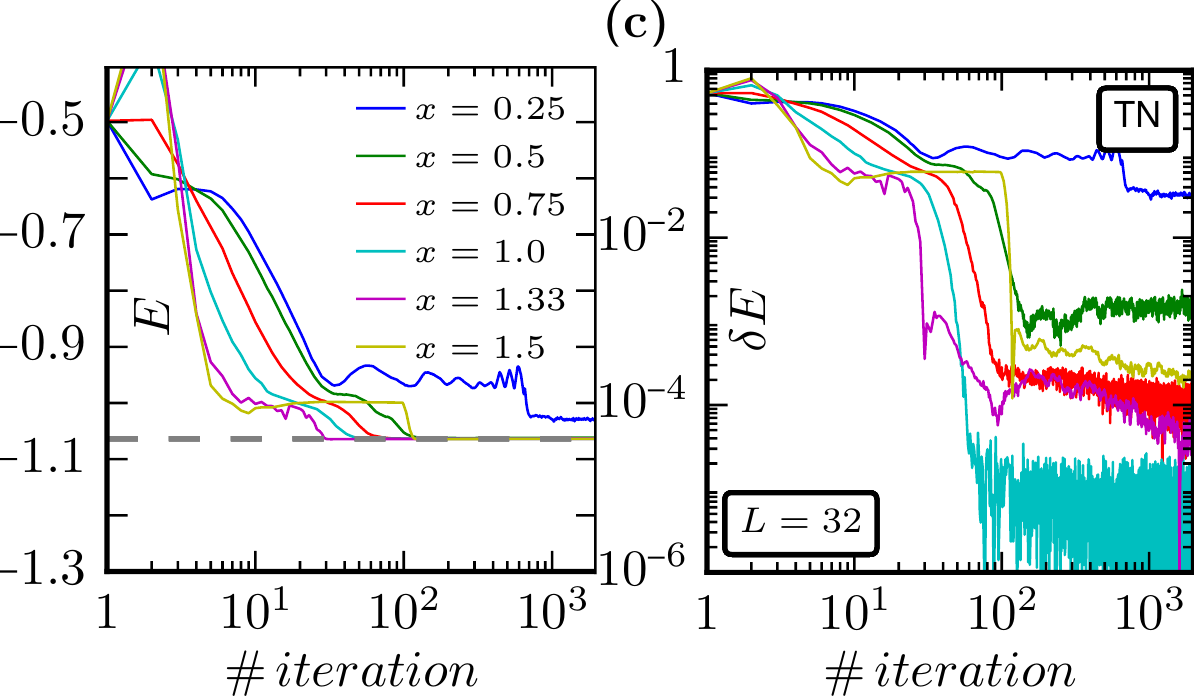} 		
 		\caption{Similar results as in Fig.~\ref{fig:energy} obtained for different system sizes in 1D.
			Sampling was carried out using Monte Carlo with $\gamma=0.2$ in panels (a) and tensor networks algorithm
			with $\gamma=0.05$  in (b). They serve as a reference point for the results in Fig.~\ref{fig:energy}.
			In panels (c) we show the influence of incorrect inverse temperature estimation, $x=\beta/\beta_x$, on the results.
}
 		\label{fig:mcmc} 
 	\end{center}    
 \end{figure}
 
\subsection{Embedding RBM on D-Wave}
Unfortunately, a RBM \emph{cannot} be directly placed on the D-Wave chip due to limited (sparse) connectivity between qubits~\cite{Benedetti17}. However, this problem can be circumvented using suitable embedding~\cite{Adachi15}. The idea
is to emulate a single neuron using available (local) connections between physical qubits on the chip. To this end, a 
strong ferromagnetic couplings is set between the latter qubits. 
We stress that even a proper embedding can break during the annealing. However, for small enough RBM's weights the 
frequency at which they do break should not be too high (in practice $\sim 0.2$). In that case, the \emph{majority vote} 
or a similar method can be invoked to correct the sample~\cite{Lidar14}.

Figure~\ref{fig:RMB}{\color{blue} (b)} shows a chimera graph with 4 unit cells. Each unit cell has $8$ qubits -- with full connectivity between the horizontal and vertical ones and can represent as
many neurons. In order to construct, for instance, a RBM with 8 visible and 8 hidden neurons, $4$ unit cells with suitable qubits ``glued" together are necessary. That amounts to $32$ physical qubits. In this embedding, all qubits connected vertically (horizontally) represent a visible (hidden) neurons~\cite{Adachi15}. Consequently, the maximum number of  neurons that the chimera graph $C_n$, consisting of $n\times n\times8$  qubits,  can represent is $L_{\rm max} = 8n$.
For example, all $2048$ qubits on the $2000$Q chips can be utilized 
to build \emph{e.g.} a RBM with 64 visible and 64 hidden neurons.
\section{Results} 
%
 \begin{figure}[t!]
 	\begin{center}
 		\includegraphics[width=\columnwidth]{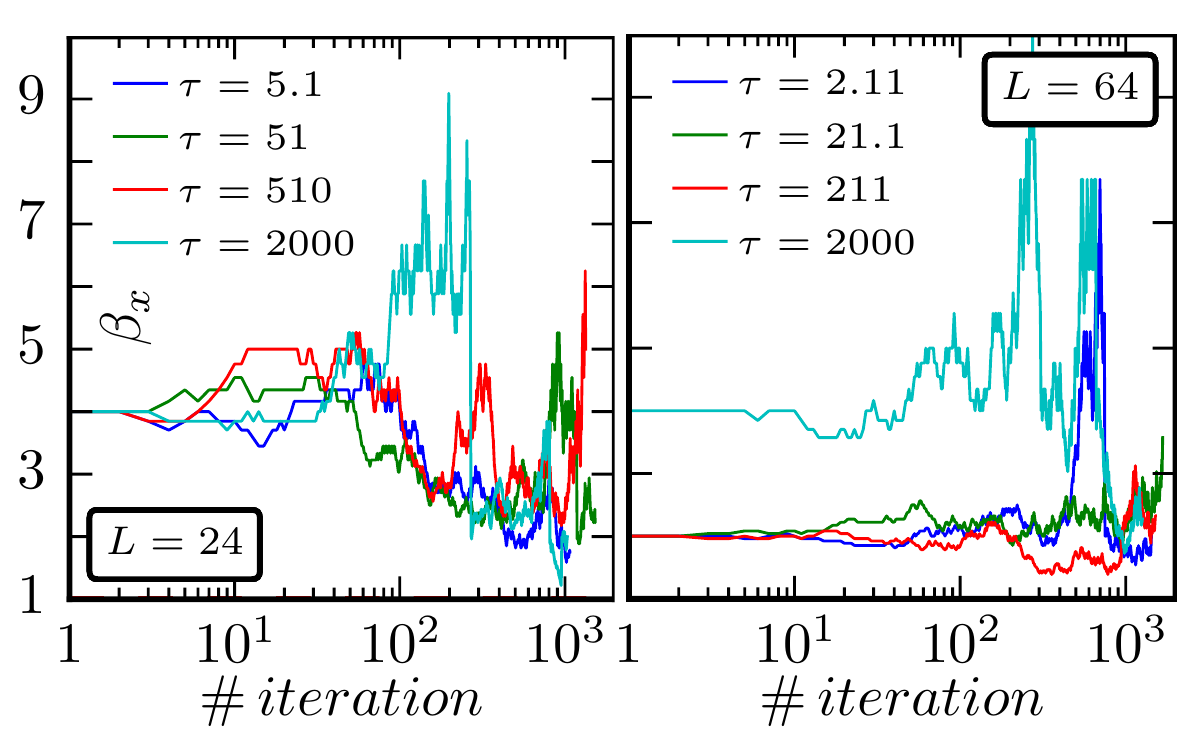}
 		\caption{
 			The inverse temperature $\beta_x$ \emph{estimated} during the learning stage. See Fig.~\ref{fig:energy} for comparison and text for discussion.
 		}
 		\label{fig:temp} 
 	\end{center}    
 \end{figure}
%
For the sake of simplicity and without loss of generality we only consider RBMs with the same number of hidden 
and visible neurons, \emph{i.e.} $M=N$ ($\alpha=1$). A classical Metropolis-Hastings sampling technique has no problems 
finding the ground state. The relative error of the solution, $\delta E = |(E-E_{\rm exact})/E_{\rm exact}|$,
%
%
is of the order of $10^{-4}$, see Fig.~\ref{fig:mcmc}{\color{blue} (a)}. The same conclusion is reached using a more sophisticated
sampling technique based on tensor networks algorithms, see Fig.~\ref{fig:mcmc}{\color{blue} (b)}. To that end we used a simpler -- matrix product states based -- variant of the
procedures described in Ref.~\cite{OTN}.
 
We collect the results which were obtained running the \emph{hybrid} algorithm in Fig.~\ref{fig:energy}. For sampling, we used two generations of D-Wave annealers: $2000$Q and DW$2$X. As one can see in Fig.~\ref{fig:energy}, both of them were capable of finding the correct ground state energy. The solutions reached are, nonetheless, less accurate with $\delta E$ of the order of $10^{-3}$---$10^{-2}$. This can be expected from the real physical device which is prone to errors~\cite{Gardas17}. One can also expect those results to improve with each new generation of quantum computers.  

There are many factors that can contribute to the errors and limited precision~\cite{Gardas17}. To mitigate some of them 
the D-Wave solver offers post-processing optimization options. The idea is to bring $p(\bfm \sigma)$ to the Boltzmann
distribution~(\ref{eq:Hf}) as close as possible, ideally at some predefined inverse temperature $\beta$. However, in our minimalistic approach we did not use any of those options. Instead, we allowed the algorithm to change the initial inverse temperature so that it could converge to the correct solution, see Fig.~\ref{fig:temp}. To that end we randomly increased or decreased the effective temperature $\beta_{\rm x}$ when the energy between subsequent iterations was growing.  
Given the lack of any comprehensive theory explaining how D-Wave annealers work, this approach seems optimal for the current purpose. 
The idea can be further motivated by numerical simulations. Fig.~\ref{fig:mcmc}{\color{blue} (c)} shows the robustness of the algorithm against variability of
$\beta_{\rm x}$. Surprisingly, the correct solution can still be reached despite incorrect estimations of the inverse temperature.

\section{Concluding remarks}
In this article we argued that despite their limited capabilities, the existing annealers can be harnessed to simulate many
body quantum systems. In our simple model a restricted Boltzmann machine was used to represent the wave function of
the transverse field quantum Ising model. Next, we show how this neural network can be trained with the help of a D-Wave annealer to find the ground state energy. The maximum system sizes that we were able to embed were restricted to $L=64$ (requiring $2048$ qubits) for $2000$Q chip and $L=24$ (requiring $\sim 800$ qubits) for its predecessor DW$2$X. This approach is nonetheless fully scalable.

As a final note, we stress that a neural network trained with an imperfect quantum annealer should, to some extent, 
reflect on the errors that are generated during the annealing~\cite{Gardas17}. This means that $\bfm w$ found by a 
faulty quantum sampler will \emph{not} produce correct results with a different sampler. This,
on the other hand, allows one to test and possibly calibrate quantum annealers against errors.

\begin{acknowledgements}
\paragraph*{Acknowledgements.} We appreciate discussions with Andy Mason, Edward Dahl and Sheir Yarkoni of D-Wave Systems. 
This work was supported by Narodowe Centrum Nauki under projects 2016/20/S/ST2/00152 (BG), 2016/23/B/ST3/00830 (JD) and funded by the National Science Centre, Poland under QuantERA programme 2017/25/Z/ST2/03028 (MMR). MMR acknowledges receiving Google Faculty Research Award 2017. This research was supported in part by PL-Grid Infrastructure.
\end{acknowledgements}

%

\end{document}